\documentclass[pra,aps,twocolumn,floatfix,showpacs]{revtex4-1}
\usepackage{graphicx,amsmath,amssymb,times}

\begin{document}
\title{Artificial gauge fields for the Bose-Hubbard model on a checkerboard \\ superlattice and extended Bose-Hubbard model}
\author{M. Iskin}
\affiliation{Department of Physics, Ko\c c University, Rumelifeneri Yolu, 34450 Sariyer, Istanbul, Turkey}
\date{\today}

\begin{abstract}We study the effects of an artificial gauge field on the ground-state phases 
of the Bose-Hubbard model on a checkerboard superlattice in two dimensions, 
including the superfluid phase and the Mott and alternating Mott insulators.
First, we discuss the single-particle Hofstadter problem, and show 
that the presence of a checkerboard superlattice gives rise to a 
magnetic flux-independent energy gap in the excitation spectrum. 
Then, we consider the many-particle problem, and derive an analytical
mean-field expression for the superfluid-Mott and 
superfluid--alternating-Mott insulator phase transition boundaries. 
Finally, since the phase diagram of the Bose-Hubbard model 
on a checkerboard superlattice is in many ways similar to that of 
the extended Bose-Hubbard model, we comment on the effects of magnetic 
field on the latter model, and derive an analytical mean-field 
expression for the superfluid-insulator phase transition boundaries as well.
\end{abstract}

\maketitle

\section{Introduction}
\label{sec:intro}

In the usual Bose-Hubbard model~\cite{bloch08}, competition 
between the kinetic and potential energy terms leads to two phases: 
a Mott insulator (MI) when the kinetic energy is much smaller 
than the potential energy and a superfluid (SF) otherwise. 
The incompressible MI phase has an excitation gap so that the 
incoherent bosons are localized, and that a slight change in 
the chemical potential does not change the number of bosons on 
a particular lattice site. The compressible SF phase, however, is 
gapless, and the coherent bosons are delocalized over the entire 
lattice. 

Recent advances with ultracold bosonic atoms loaded into optical lattices have 
made it possible to simulate Bose-Hubbard type many-particle 
Hamiltonians in a tunable setting. For instance, the ability to
control on-site boson-boson interactions has paved the way 
for observing SF and MI phases as well as the 
transition between the two~\cite{greiner02,bloch08}. 
In addition, a new technique has recently been developed 
that allowed the production of fictitious magnetic fields 
which can couple to neutral bosonic atoms~\cite{lin09a,lin09b}. 
These fictitious magnetic fields are produced through an all 
optical Raman process, couple to a fictitious charge, but produce
real effects like the creation of vortices in the SF state 
of bosons. Such an ability to control the strength of the
fictitious magnetic fields combined with the ability to control 
the strength of the interparticle interactions may allow 
exploration of new phenomena in the near 
future~\cite{niemeyer99,oktel07,umucalilar07,goldbaum09,sinha10,duric,powell,tieleman}.

In contrast to its simplicity, the Bose-Hubbard model is not 
exactly solvable even in one dimension. Therefore, it is desirable 
to have a much simpler toy-model which exhibits all the salient 
properties of the Bose-Hubbard model, while also being more amenable 
to analytical treatment. One of the most prominent candidates 
is the hardcore Bose-Hubbard model on a checkerboard superlattice, 
for which the existence of SF and MI phases at half filling in three 
dimensions~\cite{aizenman04}, as well as a direct transition 
between the two~\cite{hen09} have rigorously been shown. 
In addition, this model and its correlation functions are exactly 
solvable in one dimension~\cite{rousseau06}, thanks to the existence of 
a mapping between the hardcore bosons and noninteracting fermions.
In the absence of a magnetic field, we have recently analyzed the 
ground-state phase diagram of this model in one, two and three 
dimensions using mean-field approximation and strong-coupling 
expansion, and compared them with the numerically exact results 
obtained from the stochastic series expansion algorithm followed 
by finite-size scaling~\cite{hen10}. Given that checkerboard 
superlattices have already been realized~\cite{sebby06} 
using multiple wavelength laser beams, 
we extend previous works in two important directions. 
First, we relax the hardcore constraint and study the 
ground-state phase diagram of the softcore Bose-Hubbard model 
on a checkerboard superlattice, via mean-field decoupling
approximation. Second, we study the effects of uniform magnetic 
field on the ground-state phase diagram in two dimensions.

The rest of this paper is organized as follows. 
First, we review the model at hand in Sec.~\ref{sec:ham}, and 
present a qualitative description of its phase diagram. 
Then, we study the single-particle Hofstadter problem in 
Sec.~\ref{sec:single}, and show that the presence of a 
checkerboard superlattice gives rise to a magnetic flux-independent energy 
gap in the excitation spectrum. The many-particle problem is
discussed in Sec.~\ref{sec:many}, where we derive analytical 
expressions for the SF-insulator phase transition boundaries 
within the mean-field decoupling approximation. Then, we give a 
brief discussion and summary of our results in Sec.~\ref{sec:conc}. 
Finally, we conclude the paper with Appendix A, where an 
analytical mean-field expression for the SF-insulator phase 
transition boundaries is derived for the extended Bose-Hubbard model.

\section{Hamiltonian}
\label{sec:ham}

In this paper we study the effects of magnetic field on the 
Bose-Hubbard model on a checkerboard superlattice. For this purpose, 
we consider a two-dimensional square lattice described by the 
Hamiltonian
\begin{align}
H = &-\sum_{ij} t_{ij} e^{i\theta_{ij}} a_i^\dagger a_j + \frac{U}{2} \sum_i \widehat{n}_i (\widehat{n}_i-1) \nonumber \\  
&- C \sum_{i} (-1)^{\sigma_i} \widehat{n}_i - \mu \sum_i \widehat{n}_i,
\label{eqn:hamiltonianij} 
\end{align}
where $a_i^\dagger$ ($a_i$) creates (annihilates) a boson 
on site $i$ and $\widehat{n}_i = a_i^\dagger a_i$ is the on-site boson 
number operator. 
The hopping matrix $t_{ij}$ is assumed to connect two nearest-neighbor lattice 
sites ($t_{ij} = t$ for $i$ and $j$ nearest neighbors and $0$ otherwise) 
belonging to different sublattices, \textit{e.g.} the even sublattice 
$A$ and the odd one $B$, $U \ge 0$ is the strength of the on-site boson-boson 
repulsion, $\mu$ is the chemical potential, and
$C \ge 0$ is the amplitude of the alternating checkerboard superlattice 
potential such that $\sigma_i = 0$ $(1)$ on sublattice $A$ ($B$).
The phase factor
$
\theta_{ij} = (1/\phi_0) \int_i^j \mathbf{A_0}(\mathbf{r}) \cdot d\mathbf{r}
$
takes into account the effects of a uniform magnetic 
field that is applied perpendicular to the lattice, 
where $\mathbf{A_0}(\mathbf{r})$ is the vector potential and 
$\phi_0 = h c/e$ is the magnetic flux quantum. All of our results 
recover the nonmagnetic ones when the magnetic flux tends to 0.

Let us first analyze the atomic ($t = 0$) limit of this Hamiltonian
at zero-temperature. In this limit, there is no 
kinetic term, and the boson number operator $\widehat{n}_i$ commutes 
with the Hamiltonian, so every lattice site is occupied by a fixed number 
$n_i = \langle \widehat{n}_i \rangle$ of bosons.
Here, $\langle \dots \rangle$ is the thermal average, and the average boson 
occupancy $n_i$ is chosen to minimize the ground-state energy for a given $\mu$.
In particular, when $C = 0$, this model is translationally invariant, 
and the ground-state boson occupancy is the same for all sites.

For instance, in the hardcore boson ($U \to \infty$) limit~\cite{hen10}, 
while the lattice is completely empty 
for $\mu < 0$ and the minimal energy configuration corresponds to a 
vacuum of particles or a hole band insulator 
(VP, since $n_i = 0$ for all $i$), 
it is completely full for $\mu > 0$ and the minimal energy 
configuration corresponds to a vacuum of holes or a particle band insulator
(VH, since $n_i = 1$ for all $i$).
The ground-state energy of these phases is degenerate at $\mu = 0$. 
However, when $C > 0$, the ground state has an additional half-filled 
insulating (incompressible) phase characterized by a crystalline 
order in the form of staggered boson densities. For the nearest neighbor
lattice sites $i$ and $j$, 
$\langle \widehat{n}_i \rangle = n_A = 1$ for the sublattice $A$ and 
$\langle \widehat{n}_j \rangle = n_B = 0$ for the sublattice $B$. 
This phase resides in the region $|\mu| < C$, and it is sandwiched 
between the VP and VH phases.
Since the checkerboard superlattice breaks the translational invariance
of the lattice, it directly causes such an alternating density pattern. 
For this reason, this phase is often called a MI~\cite{hen09,hen10}, 
to distinguish it from a true charge-density-wave (CDW) phase, for which 
the translational invariance is broken spontaneously due for instance 
to the presence of nearest-neighbor interactions. (See Appendix A).
In this paper, for simplicity we call this alternating density 
pattern an alternating MI (AMI) with $(1,0)$ fillings to prevent confusion.

On one hand, for softcore bosons with $U > 2C \ne 0$, 
the ground state alternates between the AMI and MI phases as a 
function of increasing $\mu$, where the chemical potential widths 
of AMI and MI lobes are $2C$ and $U-2C$, respectively (see Fig.~\ref{fig:softpd}).
On the other hand, for softcore bosons with $2C > U$, 
the ground state has only AMI insulators. For instance, when
$2U > 2C > U$, the ground state 
   is a VP with $(0, 0)$ fillings for $\mu \le -C$; 
it is an AMI with $(1,0)$ fillings for $\mu$ between $-C$ and $U-C$; 
it is an AMI with $(2,0)$ fillings for $\mu$ between  $U-C$ and $C$; 
it is an AMI with $(2,1)$ fillings for $\mu$ between  $C$ and $2U-C$; 
it is an AMI with $(3,1)$ fillings for $\mu$ between $2U-C$ and $U+C$, and so on.
As $t$ increases, the range of $\mu$ about which the ground state 
is insulating decreases, and the MI and AMI phases disappear at 
a critical value of $t$, beyond which the system becomes
compressible. We note that, unlike the compressible SF phase of 
the usual Bose-Hubbard model, the compressible phase in this model 
is more like a supersolid (SS), where the SF and AMI orders coexist
even for arbitrarily small $C$.
However, since the checkerboard superlattice breaks the 
translational invariance of the lattice, we call the compressible 
phase of this model a SF to distinguish it from a true SS
for which the translational invariance is broken spontaneously.
Having discussed the atomic limit, let us now discuss the 
noninteracting single-particle energy spectrum when $t \ne 0$.

\section{Single-particle problem}
\label{sec:single}

In the absence of a checkerboard superlattice, i.e. when $C = 0$, 
the single-particle excitation spectrum is the usual Hofstadter 
butterfly~\cite{hofstadter76}, and here we generalize it to $C \ne 0$. 
For this purpose, we choose the Landau gauge for the vector potential 
[$\mathbf{A_0}(\mathbf{r}) \equiv (0,B_0 x,0)$], 
which leads to a uniform $\mathbf{B_0}$ field in the $z$ direction,
and the strength of the magnetic field $B_0$ is related to the 
magnetic flux $\Phi$ via $\Phi = B_0 \ell^2$.
Denoting the coordinates of lattice sites by $i \equiv (x = n\ell, y = m\ell)$, 
this gauge simply implies $\theta_{ij} = 0$ for hoppings along 
the $x$ direction, i.e. between $(n, m)$ and $(n \pm 1, m)$;  
and $\theta_{ij} = \pm 2\pi \phi n$ for links along the $y$ direction, 
i.e. between $(n, m)$ and $(n, m \pm 1)$, where $\phi = \Phi/(2\pi \phi_0)$.
In the Landau gauge, the Hamiltonian given in Eq.~(\ref{eqn:hamiltonianij}) 
can be written as
\begin{align}
\label{eqn:hamiltoniannm}
H_{sp} = -t\sum_{nm} & \left( a_{nm}^\dagger a_{n+1,m} + e^{i 2\pi\phi n} a_{nm}^\dagger a_{n,m+1} + H.c.\right) \nonumber \\
&- C \sum_{nm} (-1)^{n+m} a_{nm}^\dagger a_{nm},
\end{align}
where $n+m$ is even (odd) for sublattice $A$ ($B$). Here, we set
$U = 0$ and $\mu = 0$ for the single-particle problem.

\begin{figure} [htb]
\centerline{\scalebox{0.45}{\includegraphics{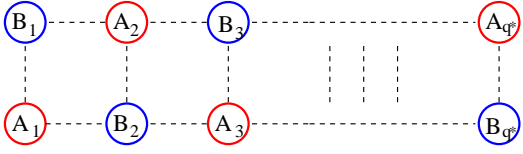}}}
\caption{\label{fig:supercell} (Color online)
The $2 \times q^*$ supercell is shown where $q^* = q$ $(2q)$ for even (odd) 
$q$ values. 
}
\end{figure}

In the Landau gauge, taking $\phi = p/q$, where $p$ and $q$ are
integers with no common factor, while the Hamiltonian given 
in Eq.~(\ref{eqn:hamiltoniannm}) maintains its checkerboard 
translational invariance in the $y$ direction, i.e. it remains 
the same under 2 steps ($m \to m+2$), it requires $q^*$ 
steps for translational invariance in the $x$ direction. 
For the square lattice considered, the period $q^* = q$ $(2q)$ for 
even (odd) $q$ values. Therefore, the Bloch theorem tells us that the 
1st magnetic Brillouin zone is determined by
$
-\pi/2 \le k_y \ell \le \pi/2
$
and
$
-\pi/q^* \le k_x \ell \le \pi/q^*.
$
This increased periodicity motivates us to work with a supercell of 
$2 \times q^*$ sites (in $x$ and $y$ directions, respectively) as shown 
in Fig.~\ref{fig:supercell}.

The single-particle excitation spectrum is determined by solving the 
Schr\"odinger equation $H_{sc} \Psi_{sc} = E_C(\phi) \Psi_{sc}$ 
for all $\mathbf{k}$ values in the 1st magnetic Brillouin zone.
Choosing the wavefunction, 
$
\Psi_{sc} = ( \psi^A_1, \psi^B_1, \psi^B_2, \psi^A_2, \psi^A_3, \psi^B_3, \dots, \psi^B_{q^*}, \psi^A_{q^*} )^T,
$
where $\psi_n^{A(B)}$ denotes the $n$th site of sublattice $A$ ($B$),
the $2q^* \times 2q*$ matrix
\begin{widetext}
%
\begin{equation}
\label{eqn:hamiltoniansc}
H_{sc} = 
\begin{bmatrix}
-C & a_1 & -te^{-ik_x\ell} & 0 & 0 & 0 & . & 0 & -te^{ik_x\ell} & 0 \\
a_1 & +C & 0 & -te^{-ik_x\ell} & 0 & 0 & . & 0 & 0 & -te^{ik_x\ell} \\
-te^{ik_x\ell} & 0 & +C & a_2 & -te^{-ik_x\ell} & 0 & . & 0 & 0 & 0 \\
0 & -te^{ik_x\ell} & a_2  & -C & 0 & -te^{-ik_x\ell} & . & 0 & 0 & 0 \\
0 & 0 & -te^{ik_x\ell} & 0 & -C & a_3 & . & 0 & 0 & 0 \\
0 & 0 & 0 & -te^{ik_x\ell} & a_3 & +C & . & . & 0 & 0 \\
. & . & . & . & . & . & . & . & . & . \\
0 & 0 & 0 & 0 & 0 & . & . & . & . & -te^{-ik_x\ell} \\
-te^{-ik_x\ell}  & 0 & 0 & 0 & 0 & 0 & . & . & +C & a_{q^*} \\
0 & -te^{-ik_x\ell} & 0 & 0 & 0 & 0 & . & -te^{ik_x\ell} & a_{q^*} & -C
\end{bmatrix}
\end{equation}
\end{widetext}
describes the supercell with periodic boundary (Bloch) conditions.
Here, $a_n = -2t\cos(k_y\ell + 2\pi n p/q)$.
Equation~(\ref{eqn:hamiltoniansc}) is the generalization of the 
Hofstadter problem to the case of a checkerboard superlattice, 
and it reduces to the usual result when $C = 0$~\cite{hofstadter76}. 

\begin{figure} [htb]
\centerline{\scalebox{0.4}{\includegraphics{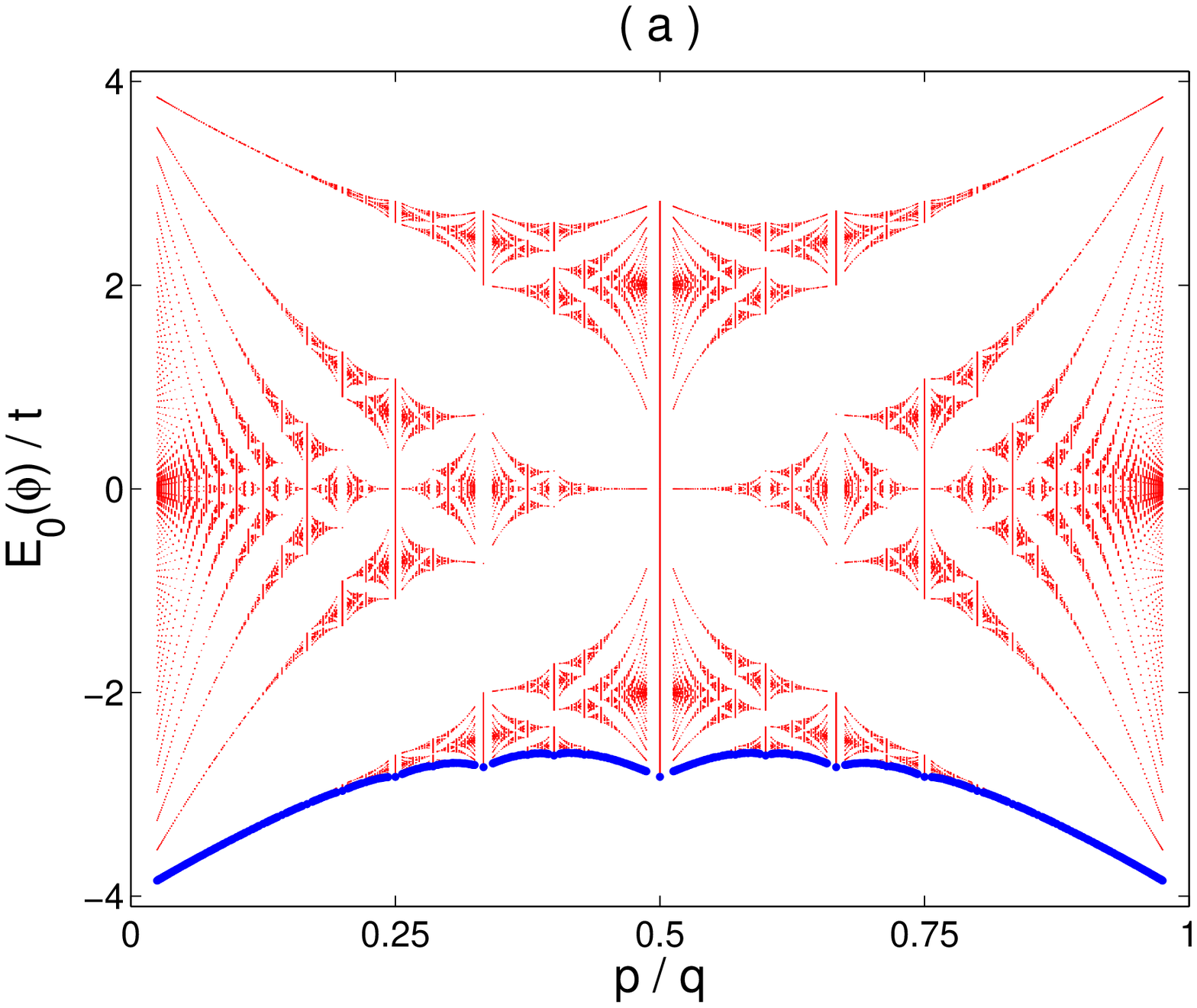}}}
\centerline{\scalebox{0.4}{\includegraphics{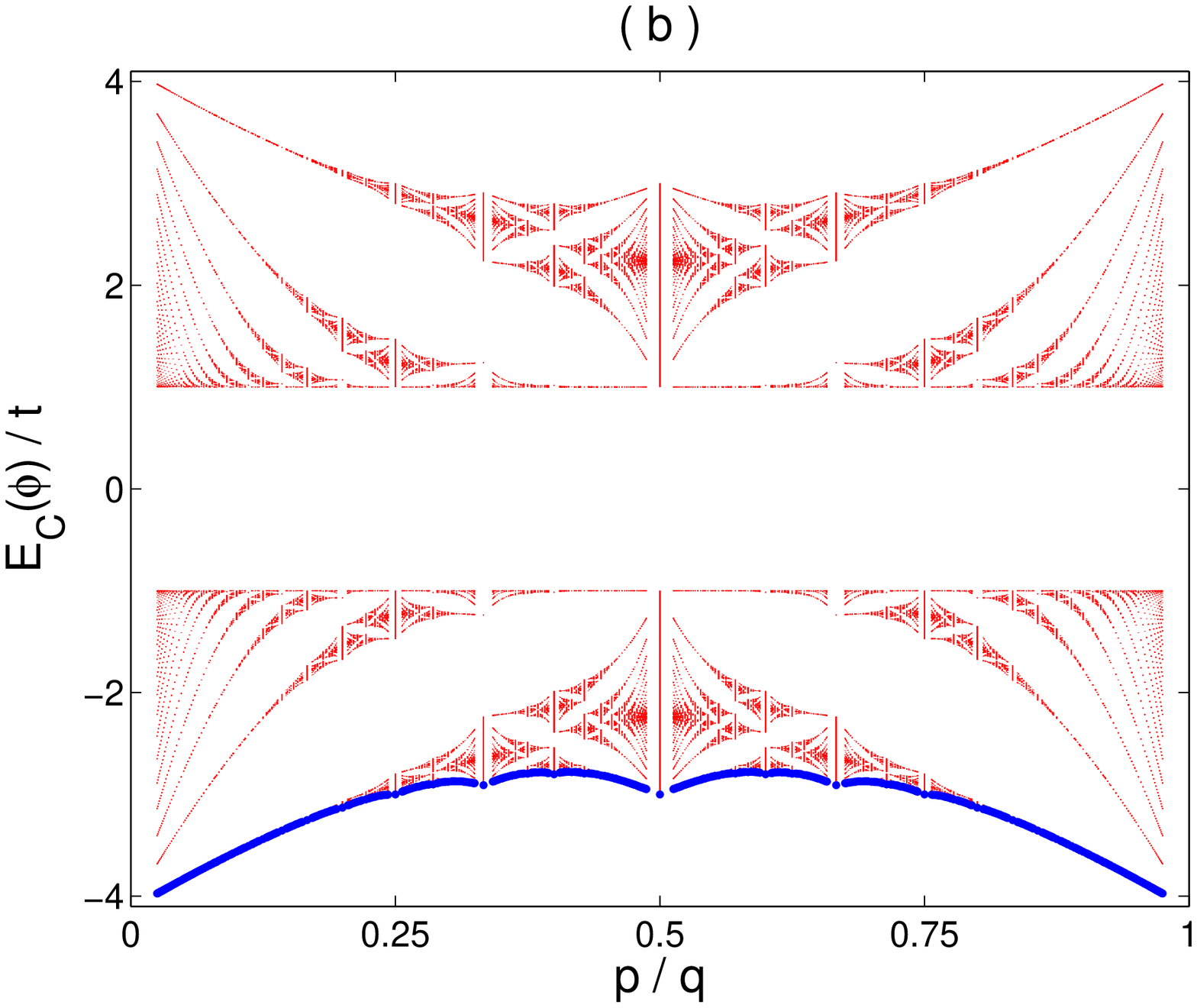}}}
\caption{\label{fig:spectrum} (Color online)
The single-particle excitation spectrum $E_C(\phi)$ of the Hamiltonian 
given in Eq.~(\ref{eqn:hamiltoniansc}) (in units of $t)$ 
is shown as a function of the magnetic flux $\phi = \Phi/(2\pi \phi_0) =  p/q$ 
for two values of $C$: (a) $C = 0$ and (b) $C = t$. Note that a 
flux-independent energy gap opens in (b), which exactly equals to $2C$.
}
\end{figure}

In Fig.~\ref{fig:spectrum}, we show the single-particle 
excitation spectrum $E_C(\phi)$ as a function of the magnetic flux
$\phi = p/q$ for two values of $C$: (a) $C = 0$ and (b) $C = t$.
In both figures, $E_C(\phi)$ is shown to be symmetric around $p/q = 1/2$,
i.e. $E_C(\phi) = E_C(1 - \phi)$. This is simply because the
magnetic flux values that add up to $2\pi \phi_0$ are equivalent,
e.g. $\phi = 0$ and $\phi = 1$. This also means that the
maximal magnetic field $B_0$ that can be applied corresponds 
to $p/q = 1/2$. A list of minimal single-particle excitation 
energies $\epsilon(\phi) = \min_{\mathbf{k}} E_0(\phi)/t$ of the usual Hofstadter 
butterfly is also given for a number of $p/q$ values 
in Table.~\ref{table:emin}. 

In contrast to the presence of zero-energy excitations for all
possible $\phi$ values in Fig.~\ref{fig:spectrum}(a),
the most important difference in Fig.~\ref{fig:spectrum}(b)
is the presence of a flux-independent energy gap. 
Our numerical calculations show that the energy gap is exactly $2C$. 
Since the on-site energy difference between the sublattice A and B is
2C, this result is not so surprising at least in the $\phi \to 0$ 
limit. In an earlier work~\cite{hen10}, in the absence of a magnetic field, 
we showed that the single-particle excitation spectrum becomes 
$
E_C(0) = \pm \sqrt{C^2 + E_0^2(0)},
$ 
where 
$
E_0(0) = - 2t\cos(k_x\ell) - 2t\cos(k_y\ell)
$ 
is the usual single-particle excitation spectrum. Similar to the 
nonmagnetic case, our numerical calculations show that
\begin{equation}
E_C(\phi) = \pm \sqrt{C^2 + E_0^2(\phi)}
\label{eqn:EC}
\end{equation}
holds exactly in the presence of a magnetic field, where 
$E_0(\phi) = E_{C = 0}(\phi)$ 
is the usual single-particle excitation spectrum of the 
Hofstadter butterfly. 
Having discussed the single-particle problem, now we are ready
to analyze the competition between the kinetic and potential 
energy terms of the many-particle Hamiltonian when $t \ne 0$. 

\begin{center}
\begin{table} [htb]
\caption{\label{table:emin} 
A list of minimal single-particle excitation energies 
$\epsilon(\phi) = \min_{\mathbf{k}} E_0(\phi)/t$ of the usual Hofstadter butterfly ($C = 0$)
is given for a number of magnetic flux $\phi = \Phi / (2\pi\phi_0) = p/q$ values.
}
\begin{tabular}{cc|cc|cc|cc}
\hline \hline
$p/q$  &  $\epsilon(\phi)$  &  $p/q$  &  $\epsilon(\phi)$  &  
$p/q$  &  $\epsilon(\phi)$  &  $p/q$  &  $\epsilon(\phi)$ \\
\hline
1/2  & -2.828  &  2/3	&  -2.732  &  3/4   &  -2.828  &  4/5   &  -2.966  \\
1/3  & -2.732  &  2/5   &  -2.618  &  3/5   &  -2.618  &  4/7   &  -2.611  \\
1/4  & -2.828  &  2/7   &  -2.725  &  3/7   &  -2.611  &  4/9   &  -2.630  \\
1/5  & -2.966  &  2/9   &  -2.881  &  3/8   &  -2.613  &  4/11  &  -2.626  \\
1/6  & -3.096  &  2/11  &  -3.028  &  3/10  &  -2.698  &  4/13  &  -2.692  \\
1/7  & -3.203  &  2/13  &  -3.151  &  3/11  &  -2.746  &  4/15  &  -2.761  \\
1/8  & -3.291  &  2/15  &  -3.249  &  3/13  &  -2.854  &  4/17  &  -2.842  \\
1/9  & -3.362  &  2/17  &  -3.328  &  3/14  &  -2.906  &  4/19  &  -2.920  \\
1/10 & -3.420  &  2/19  &  -3.392  &  3/16  &  -3.005  &  4/21  &  -2.994  \\
1/11 & -3.469  &  2/21  &  -3.445  &  3/17  &  -3.050  &  4/23  &  -3.061  \\
1/12 & -3.510  &  2/23  &  -3.490  &  3/19  &  -3.132  &  4/25  &  -3.123  \\
1/13 & -3.545  &  2/25  &  -3.528  &  3/20  &  -3.168  &  4/27  &  -3.177  \\
1/14 & -3.576  &  2/27  &  -3.561  &  3/22  &  -3.234  &  4/29  &  -3.226  \\
1/15 & -3.602  &  2/29  &  -3.590  &  3/23  &  -3.263  &  4/31  &  -3.263  \\
\hline \hline
\end{tabular}
\end{table}
\end{center}
\section{Many-particle problem}
\label{sec:many}

For the many-particle problem, we add the boson-boson interaction and 
chemical potential terms to Eq.~(\ref{eqn:hamiltoniannm}), and obtain
\begin{align}
\label{eqn:hamiltonianmp}
H_{mp} = &-t\sum_{nm} \left( a_{nm}^\dagger a_{n+1,m} + e^{i 2\pi\phi n} a_{nm}^\dagger a_{n,m+1} + H.c.\right) \nonumber \\
&+ \frac{U}{2} \sum_{nm} a_{nm}^\dagger a_{nm} (a_{nm}^\dagger a_{nm} - 1) \nonumber \\
&- C \sum_{nm} (-1)^{n+m} a_{nm}^\dagger a_{nm} - \mu  \sum_{nm} a_{nm}^\dagger a_{nm}.
\end{align}
For illustrative purposes, let us first study the ground-state 
phase diagram of this model for the hardcore bosons, for which the 
calculation becomes considerably simpler compared to the softcore case,
but yet nontrivial.

\subsection{Hardcore bosons}
\label{sec:hard}

The hardcore boson operators satisfy the constraint 
$a^{\dagger 2}_{nm} = a^2_{nm} = 0$, which prohibit multiple 
occupancy of lattice sites, as dictated by the 
infinitely large on-site boson-boson repulsion ($U \rightarrow \infty$). 
In this limit, the many-particle Hamiltonian given in Eq.~(\ref{eqn:hamiltonianmp}) 
becomes invariant under the transformation $a_{nm} \to a_{n\pm 1,m}^\dagger$ or 
$a_{nm} \to a_{n,m\pm 1}^\dagger$, which corresponds to a 
shift of one lattice site in $x$ or $y$ direction. This symmetry 
operation, which can be immediately read off from the Hamiltonian, 
corresponds to a particle-hole exchange combined with swapping 
$A$ and $B$ sublattices. It leads to a $\mu \to - \mu$ symmetry, 
i.e. the phase diagram is symmetric around $\mu = 0$.

It turns out that the exact SF-VP and SF-VH phase transition 
boundaries can be easily obtained analytically. 
The simplest argument leading to this conclusion stems from 
the fact that this boundary is determined by the addition 
of a single particle (hole) to the completely-empty (-filled) lattice.
It can then be argued that whether one is dealing with hardcore 
bosons or noninteracting spinless fermions makes no difference, 
as the particle statistics plays no role. 
This further means that one needs only to diagonalize the 
single-particle Hamiltonian and find the energy difference between 
the completely-empty (-filled) lattice and the state with one 
particle (hole). The single-particle spectrum is already given 
in Eq.~(\ref{eqn:EC}), and this procedure leads to
\begin{align}
\mu = \pm \sqrt{C^2 + \epsilon^2(\phi) t^2},
\label{eqn:mubi}
\end{align}
where $\epsilon(\phi) = \min_{\mathbf{k}} E_0(\phi)/t$ are the minimal 
single-particle excitation energies of the usual Hofstadter 
butterfly which are shown as big blue dots in Fig.~\ref{fig:spectrum}(a). 
In other words, the minus (plus) branch in Eq.~(\ref{eqn:mubi}) 
is determined by the minimal (maximal) single-particle excitation 
energies of the Hofsdtadter butterfly when $C \ne 0$. The minimal  
single-particle excitation energies are shown as big blue dots in 
Fig.~\ref{fig:spectrum}(b). We emphasize that Eq.~(\ref{eqn:mubi}) 
is exact for two-dimensional square lattices, and the minus (plus) 
sign determines the SF-VP (-VH) phase transition boundary.
Note again that Eq.~(\ref{eqn:mubi}) reduces to the known result 
for the nonmagnetic case~\cite{hen10} in the $\epsilon(\phi \to 0) = -4$ limit. 

Unlike the SF-VP and SF-VH phase boundaries, the SF-AMI phase transition 
boundary cannot be determined exactly, since the exact many-particle wave 
function for the AMI state is not known. However, this can be
achieved via the mean-field decoupling approximation. 
Within this approximation, the hopping terms in the Hamiltonian 
given in Eq.~(\ref{eqn:hamiltonianmp}) are decoupled according to,
$
a_{nm}^\dagger a_{n+1,m} = \langle a_{nm}^\dagger \rangle a_{n+1,m} 
+ a_{nm}^\dagger \langle a_{n+1,m} \rangle 
- \langle a_{nm}^\dagger \rangle \langle a_{n+1,m} \rangle,
$
where the expectation values $\varphi_{nm}(\phi) = \langle a_{nm} \rangle$ 
correspond to the mean-field SF order parameters. Note that there 
are at most $2q^*$, i.e. total number of sites in a supercell,
distinct $\varphi_{nm}$ and the SF (AMI) phase is determined by 
the nonzero (zero) value of any one (all) of them. 
In particular, there are two distinct SF order parameters 
(one for each sublattice) in the nonmagnetic case even for 
arbitrarily small $C$, as long as $C \ne 0$. 
We checked that these order parameters differ from each other 
for all parameter space via Gutzwiller ansatz calculations. 
Therefore, unlike the compressible SF phase of the usual 
Bose-Hubbard model, the SF phase in the checkerboard model is 
more like a SS, where the SF and AMI orders coexist.

Performing a second-order perturbation theory 
in $\varphi_{nm}$ around the MI phase, 
and following the usual Landau procedure for second-order phase 
transitions, i.e. minimizing the ground-state energy as a function 
of $\varphi_{nm}$, we eventually arrive at the phase 
transition boundary equation
\begin{align}
\mu = \pm \sqrt{C^2 - \epsilon^2(\phi) t^2},
\label{eqn:mumi}
\end{align}
where the plus (minus) sign corresponds to particle (hole) excitations
above the AMI phase. We note that although $\varphi_{nm}$ 
are gauge dependent, the phase transition boundary itself 
is not~\cite{oktel07,umucalilar07,sinha10}. 
Alternatively, Eq.~(\ref{eqn:mumi}) follows 
directly from the strong-coupling expansion of the ground-state 
energy of the AMI phase with respect to the hopping term~\cite{hen10}, 
where the first nontrivial hopping dependence of the 
phase transition boundary arises from the maximal eigenvalue 
of the $\mathbf{T} = \mathbf{t} \cdot \mathbf{t}$ 
matrix~\cite{noteBH}. Here, the elements of $T_{ij} = \sum_k t_{ik} t_{kj}$ 
are such that $\sum_{j} T_{ij} f_{j} = \epsilon^2(\phi) t^2 f_i$. 
Note again that Eq.~(\ref{eqn:mumi}) reduces to the known result 
for the nonmagnetic case~\cite{hen10} in the $\epsilon(\phi \to 0) = -4$ limit. 

\begin{figure} [htb]
\centerline{\scalebox{0.217}{\includegraphics{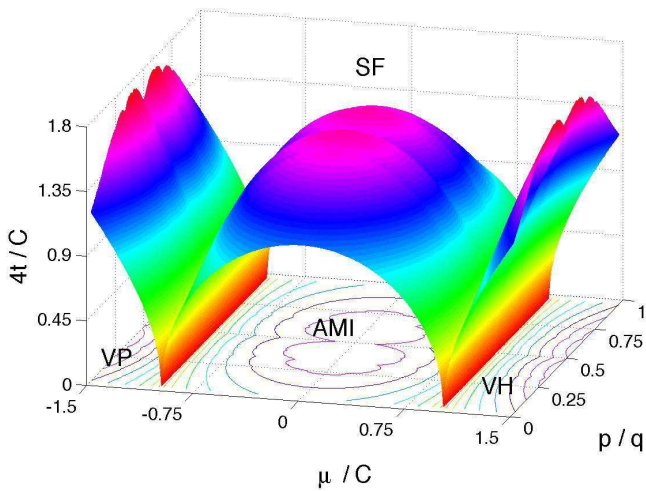}}}
\caption{\label{fig:pd} (Color online)
The ground-state phase diagram of hardcore bosons is shown as a function of 
the chemical potential $\mu$ (in units of $C$), hopping $4t$ 
(in units of $C$) and magnetic flux $\phi = \Phi / (2\pi\phi_0) = p/q$, 
as obtained from Eqs.~(\ref{eqn:mubi}) and~(\ref{eqn:mumi}). 
The phase diagram is symmetric around $\mu = 0$ and $p/q = 1/2$ 
as explained in the text, and the intriguing structure of the 
phase boundaries (which is clearly seen in the contour plot) 
is due to the minimal single-particle excitation energies 
$\epsilon(\phi)$ shown as big blue dots in Fig.~\ref{fig:spectrum}(a). 
}
\end{figure}

In Fig.~\ref{fig:pd}, we show the ground-state phase diagram 
as a function of the chemical potential $\mu$, hopping $4t$ and 
magnetic flux $\phi = p/q$, that is obtained from 
Eqs.~(\ref{eqn:mubi}) and~(\ref{eqn:mumi}). As we argued above, 
the phase diagram is symmetric around $\mu = 0$ and $p/q = 1/2$. 
The latter symmetry is in agreement with the earlier findings 
on the usual Bose-Hubbard model~\cite{niemeyer99,oktel07,umucalilar07}. 
What is more interesting is the intriguing structure of the phase
transition boundaries on the minimal single-particle excitation 
energies $\epsilon(\phi)$.
In addition, the incompressible (compressible) AMI (SF) 
phase grows (shrinks) when the magnetic field increases from 
zero, due to the localizing effects 
of the magnetic field on bosons. All of these observations 
are similar to earlier findings on the usual Bose-Hubbard 
model~\cite{niemeyer99,oktel07,umucalilar07}.

\subsection{Softcore bosons}
\label{sec:soft}

Having studied the effects of magnetic field on the hardcore bosons,
let us now analyze the ground-state phase diagram of the Hamiltonian 
given in Eq.~(\ref{eqn:hamiltonianmp}) for the softcore bosons. 
Since the exact many-particle wave functions for the AMI and MI 
states are not known, we again obtain the phase diagram via the 
mean-field decoupling approximation. 
Following the recipe given in the previous section, the SF-MI and 
SF-AMI phase transition boundaries are found to be determined by
\begin{align}
\label{eqn:soft}
\frac{1}{\epsilon^2(\phi) t^2} &= \left[ \frac{n_A+1}{U n_A - C - \mu} 
- \frac{n_A}{U (n_A-1) - C - \mu} \right] \nonumber \\
\times &\left[ \frac{n_B+1}{U n_B + C -\mu} -\frac{n_B}{U (n_B-1) + C -\mu} \right],
\end{align}
which gives a quartic equation for $\mu$. This equation is valid 
for all $C$, and $\epsilon(\phi) = \min_{\mathbf{k}} E_0(\phi)/t$ are the minimal 
single-particle excitation energies of the usual Hofstadter butterfly. 
In the hardcore ($U \to \infty$) limit, note that Eq.~(\ref{eqn:soft}) 
reduces to Eq.~(\ref{eqn:mumi}) when $n_A = 1$ and $n_B = 0$, and to
Eq.~(\ref{eqn:mubi}) when $n_A = n_B = 0$ or $n_A = n_B = 1$.
In addition, Eq.~(\ref{eqn:soft}) reduces to the known result~\cite{umucalilar07} 
for the usual Bose-Hubbard model when $n_A = n_B = n_0$ and $C = 0$, 
and it also agrees with the recent numerical calculations~\cite{buonsante04,chen10} 
in the absence of a magnetic field.
Since a simple closed form analytic solution for $\mu$ is not possible
when $C \ne 0$, we solve Eq.~(\ref{eqn:soft}) with MATHEMATICA for each 
of the AMI and MI lobes separately. 

\begin{figure} [htb]
\centerline{\scalebox{0.7}{\includegraphics{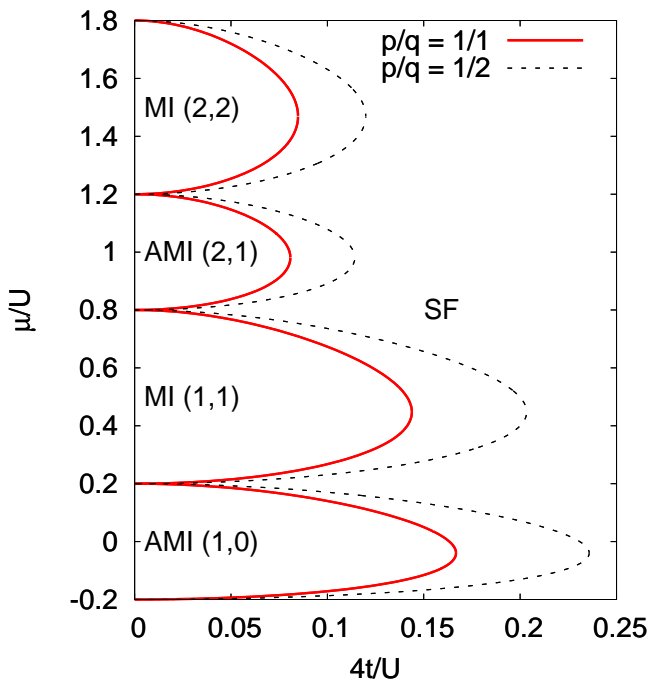}}}
\caption{\label{fig:softpd} (Color online)
The ground-state phase diagram of softcore bosons is shown as a function of 
the chemical potential $\mu$ (in units of $U$) and hopping $4t$ 
(in units of $U$), as obtained from Eq.~(\ref{eqn:soft}). 
Here, we set $C = 0.2U$, and $p/q = 1/1$ $(1/2)$
is shown as solid red (dashed black), corresponding to zero 
(maximum) magnetic field. 
The AMI and MI (SF) phases grow (shrinks) when the magnetic 
field increases from zero, due to the 
localizing effects of the magnetic field on bosons.
}
\end{figure}

In Fig.~\ref{fig:softpd}, we set $C = 0.2U$ and show the 
ground-state phase diagram as a function of the chemical 
potential $\mu$ and hopping $4t$ for $p/q = 1/1$ 
(equivalent to zero magnetic field) and for $p/q = 1/2$ 
(maximum magnetic field). 
As discussed in Sec.~\ref{sec:ham}, the ground state alternates 
between the AMI and MI phases as a function of increasing $\mu$.
The chemical potential widths of AMI and MI lobes are 
$0.4U$ and $0.6U$, respectively, but the size of the AMI (MI)
lobes increase (decrease) as a function of increasing $C/U$ 
(not shown), since a nonzero $C$ is what allowed AMI states 
to form in the first place. In addition, the incompressible 
(compressible) AMI and MI (SF) phases grow (shrinks) when the 
magnetic field increases from zero, due again to the localizing effects of the magnetic field 
on bosons. Since the phase diagram of the Bose-Hubbard model 
on a checkerboard superlattice is in many ways similar to that of 
the extended Bose-Hubbard model, we discussed the latter model 
in Appendix A.

\section{Conclusions}
\label{sec:conc}

In this paper, we studied ground-state phases of the 
Bose-Hubbard model on a checkerboard superlattice in two dimensions. 
First, we discussed the single-particle Hofstadter problem, and 
showed that the presence of a checkerboard superlattice gives rise to 
a magnetic flux-independent energy gap in the excitation spectrum. 
Then, we considered the many-particle problem, and derived analytical
mean-field expressions for the SF-MI and SF-AMI phase transition 
boundaries. We showed that the size of incompressible insulator 
phases grow when the magnetic field increases from zero, 
due to the localizing effect of the magnetic field 
on bosons. In addition, since the phase boundaries are functions 
of the minimal single-particle excitation energies, they have 
an intriguing dependence on the magnetic flux. 

We also showed that the phase diagram of the Bose-Hubbard model 
on a checkerboard superlattice is in many ways similar to that of 
the extended Bose-Hubbard model. In particular, the compressible
phase in the former model is more like a 
SS, where the SF and AMI orders coexist even for arbitrarily small $C$.
However, since the checkerboard superlattice breaks the 
translational invariance of the lattice, we call the compressible 
phase of this model a SF to distinguish it from a true SS
for which the translational invariance is broken spontaneously.
For completeness, we discussed the effects of magnetic field on 
the the extended Bose-Hubbard model as well, and derived an analytical 
mean-field expression for the SF-MI and SF-CDW phase transition 
boundaries.

In this paper, we relied on the mean-field theory which is known
to be sufficient in describing only the qualitative features 
of the phase diagram when there is no magnetic field. 
In particular, for the hardcore Bose-Hubbard model on a 
checkerboard superlattice, we have recently shown that the 
mean-field theory has a large quantitative discrepancy from 
the numerically exact Quantum Monte Carlo results especially 
in lower dimensions~\cite{hen10}. However, due to the infamous 
sign problem, such numerical calculations cannot be performed 
in the presence of a magnetic field, and therefore, it is not 
clear to us whether the mean-field theory is sufficient in this 
case or not. We hope that the intriguing structure of the phase
transition boundaries on the minimal single-particle excitation 
energies of the Hofstadter butterfly, pedicted by the mean-field 
theory, could be observed in the experiments or verified via other 
exact means in the future.

We plan to extend this work at least in one important direction. 
There is some evidence that the ground-state phase diagram of the
extended Bose-Hubbard model includes a SS phase, in dimensions higher 
than one~\cite{bruder93,parhat94,kovrizhin05,iskin11}. The localizing 
effect of magnetic field on such a phase is yet to be studied, and
it is not obvious whether SS region would grow or shrink
when the magnetic field increases from zero.
Although numerical calculations based on the Gutzwiller ansatz 
are expected to give gauge dependent results for the SS-SF phase boundary 
because of the mean-field nature of the ansatz, they would provide 
a good qualitative insight into this problem.

\section{Acknowledgments}
\label{sec:ack}

The author thanks I. Hen for correspondence.
This work is supported by the Marie Curie International Reintegration 
(Grant No. FP7-PEOPLE-IRG-2010-268239), Scientific and Technological 
Research Council of Turkey (Career Grant No. T\"{U}B$\dot{\mathrm{I}}$TAK-3501-110T839), 
and the Turkish Academy of Sciences (T\"{U}BA-GEB$\dot{\mathrm{I}}$P).

\appendix

\section{Extended Bose-Hubbard model}
\label{sec:eBH}

In many ways, the phase diagram of the Bose-Hubbard model on a 
checkerboard superlattice turned out to be similar to that of the 
extended Bose-Hubbard model. Therefore, in this appendix, we comment 
on the effects of uniform magnetic field on the insulating 
phases of the latter model. In contrast to our model where the 
translational invariance is broken due to checkerboard superlattice, 
the translational invariance is broken spontaneously in the extended 
model, leading to CDW modulations. 
The extended Bose-Hubbard Hamiltonian with the on-site ($U \ge 0$) 
and nearest-neighbor ($V \ge 0$) boson-boson repulsions can be written as
\begin{align}
\label{eqn:ebhh}
H = &- t\sum_{\langle ij \rangle} \left( e^{i\theta_{ij}} a_i^\dagger a_j + H.c. \right)
+ \frac{U}{2} \sum_i \widehat{n}_i (\widehat{n}_i-1) \nonumber \\
&+ V \sum_{\langle ij \rangle} \widehat{n}_i \widehat{n}_j -\mu \sum_i \widehat{n}_i.
\end{align}
Here, we again consider a two-dimensional square optical lattice.
For $U > 4V \ne 0$, it is well-known that the ground state 
has two types of insulating phases~\cite{bruder93,parhat94,kovrizhin05,iskin09}. 
The first one is the MI phase where, similar to the usual 
Bose-Hubbard model, the ground-state boson occupancy is the 
same for every lattice site, i.e. $\langle \widehat{n}_i \rangle = n_0$. 
The second one is the CDW phase which has crystalline order in 
the form of staggered boson occupancies, i.e.
$\langle \widehat{n}_i \rangle = n_A$ and 
$\langle \widehat{n}_j \rangle = n_B$ for $i$ and $j$ nearest neighbors. 

\begin{figure} [htb]
\centerline{\scalebox{0.7}{\includegraphics{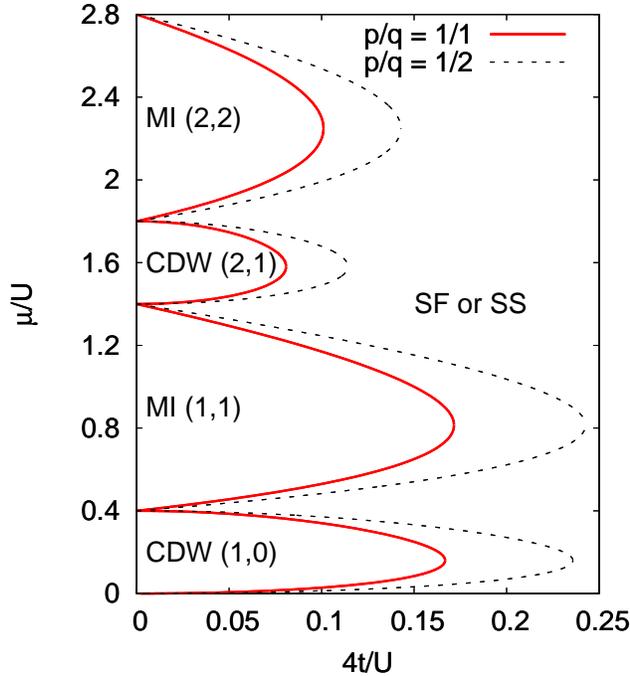}}}
\caption{\label{fig:eBHpd} (Color online)
The ground-state phase diagram of the extended Bose-Hubbard model 
is shown as a function of the chemical potential $\mu$ 
(in units of $U$) and hopping $4t$ 
(in units of $U$), as obtained from Eq.~(\ref{eqn:mf}). 
Here, we set $V = 0.1U$, and $p/q = 1/1$ $(1/2)$
is shown as solid red (dashed black), corresponding to zero 
(maximum) magnetic field. 
}
\end{figure}

When $U > 4V$, in the atomic $t = 0$ limit, the ground 
state alternates between the CDW and 
MI phases as a function of increasing $\mu$, where the chemical 
potential widths of CDW and MI lobes are $4V$ and $U$, respectively. 
As $t$ increases, the range of $\mu$ about which the ground state 
is insulating decreases, and the MI and CDW phases disappear at 
a critical value of $t$, beyond which the system becomes 
compressible (SF or SS).
On the other hand, the ground state has only CDW insulators 
when $U < 4V$. The chemical potential width of all 
CDW insulators is $U$, and the ground state is a 
CDW insulator with $(n_0, 0)$ fillings for 
$(n_0-1) U < \mu < n_0 U$. As $t$ increases, the CDW phases 
disappear at a critical value of $t$, beyond which the system 
first becomes a SS then a SF at a much larger $t$ with a very 
large region of SS phase~\cite{iskin11}. 
Within the mean-field decoupling theory, the phase transition 
boundaries are determined by
\begin{align}
\label{eqn:mf}
\frac{1}{\epsilon^2(\phi) t^2} &= \left[ \frac{n_A+1}{U n_A + 4V n_B - \mu} 
- \frac{n_A}{U (n_A-1) + 4 V n_B - \mu} \right] \nonumber \\
\times &\left[ \frac{n_B+1}{U n_B + 4V n_A -\mu} -\frac{n_B}{U (n_B-1) + 4V n_A -\mu} \right],
\end{align}
which gives a quartic equation for $\mu$. Here, 
$\epsilon(\phi) = \min_{\mathbf{k}} E_0(\phi)/t$ depends on the minimal 
single-particle excitation energies of the usual Hofstadter butterfly. 
Note that Eq.~(\ref{eqn:mf}) reduces to the known result for 
the nonmagnetic case~\cite{iskin09} in the $\epsilon(\phi \to 0) = -4$
limit, and it reduces to the known magnetic 
result~\cite{niemeyer99,oktel07,umucalilar07} when 
$n_A = n_B = n_0$ and $V = 0$. Equation~(\ref{eqn:mf}) shows that 
the MI lobes are separated by $4V$, but their shapes are independent 
of $V$ within the mean-field decoupling approximation; in particular, 
the critical points for the MI lobes are independent of $V$.

In Fig.~\ref{fig:eBHpd}, we set $V = 0.1U$ and show the 
ground-state phase diagram as a function of the chemical 
potential $\mu$ and hopping $4t$ for $p/q = 1/1$ 
(no magnetic flux) and for $p/q = 1/2$ (maximum magnetic flux). 
As discussed above, the ground state alternates 
between the CDW and MI phases as a function of increasing $\mu$.
The chemical potential widths of CDW and MI lobes are 
$0.4U$ and $U$, respectively, but the size of the CDW (MI)
lobes increase (decrease) as a function of increasing $V/U$ 
(not shown), since a nonzero $V$ is what allowed CDW states 
to form in the first place. In addition, the incompressible 
(compressible) CDW and MI (SF or SS) phases grow (shrinks) 
when the magnetic field increases from zero, due to the localizing 
effects of the magnetic field on bosons.


\begin{thebibliography}{99}
\bibitem{noteBH} This is in contrast to the usual Bose-Hubbard model where 
the first nontrivial hopping dependence of the phase transition boundary arises 
from the minimal eigenvalue of the $-\mathbf{t}$ matrix, such that 
$-\sum_{j} t_{ij} f_{j} = \epsilon(\phi) t f_i$~\cite{niemeyer99}.
\end{thebibliography}
\end{document}